\documentclass[aps,pre,reprint,twocolumn,showpacs,floatfix,nofootinbib, showkeys,longbibliography,superscriptaddress]{revtex4-2}
\usepackage{subfigure}
\usepackage{amssymb}
\usepackage{amsfonts}
\usepackage{amsmath}
\usepackage{amsthm}
\usepackage{epsfig}
\usepackage{graphicx}
\usepackage[usenames,dvipsnames]{color}
\usepackage[latin1]{inputenc}

\usepackage[hidelinks,unicode=true]{hyperref}
\hypersetup{colorlinks=true,
	linkcolor=blue,
	urlcolor=blue,
	citecolor=blue,
	pdfhighlight=/N
}

\usepackage{comment}
\usepackage[normalem]{ulem} 

\newcommand{\av}[1]{\langle{#1}\rangle}

\newcommand{\Ninf}{N_\text{inf}}
\newcommand{\Ne}{N_\text{e}}

\graphicspath{{figuras/}}

\begin{document}
\title{Simple quasistationary method for simulations of epidemic processes 
with localized states}

\author{Guiherme S. Costa }
\affiliation{Departamento de F\'{\i}sica, Universidade Federal de Vi\c{c}osa, 36570-900 Vi\c{c}osa, Minas Gerais, Brazil}

\author{Silvio C. Ferreira}
\affiliation{Departamento de F\'{\i}sica, Universidade Federal de Vi\c{c}osa, 36570-900 Vi\c{c}osa, Minas Gerais, Brazil}
\affiliation{National Institute of Science and Technology for Complex Systems, 22290-180, Rio de Janeiro, Brazil}

\begin{abstract}
Epidemic processes on random graphs or networks are marked by localization of activity that can trap the dynamics into a metastable state, confined to a subextensive  part of the network, before visiting an absorbing configuration. Quasistationary (QS) method is a  technique to deal with  absorbing states for finite sizes and has played a central role in the investigation of epidemic processes on heterogeneous networks where localization is a hallmark. The standard QS method possesses high computer and algorithmic complexity for large systems besides parameters whose choice are not systematic. However, simpler approaches, such as a reflecting boundary condition (RBC), are not able to capture the localization effects as the standard QS method does. In the present work, we propose a QS method that consists of reactivating  nodes proportionally to the time they were active along the preceding simulation. The method is compared with the standard QS and RBC methods for the susceptible-infected-susceptible model on complex networks, which is a prototype of a dynamic process with strong and localization effects. We verified that the method performs as well the as standard QS in all investigated simulations, providing the same scaling exponents, epidemic thresholds, and localized phases, thus overcoming the limitations of other simpler approaches. We also report that the present method has significant reduction of the computer and algorithmic complexity than the standard QS methods. So, this method arises as a simpler and efficient tool to analyze localization on heterogeneous structures through QS simulations.
\end{abstract}


\maketitle

\section{Introduction}

\begin{figure}[hbt]
	
	\includegraphics[width=0.8\columnwidth]{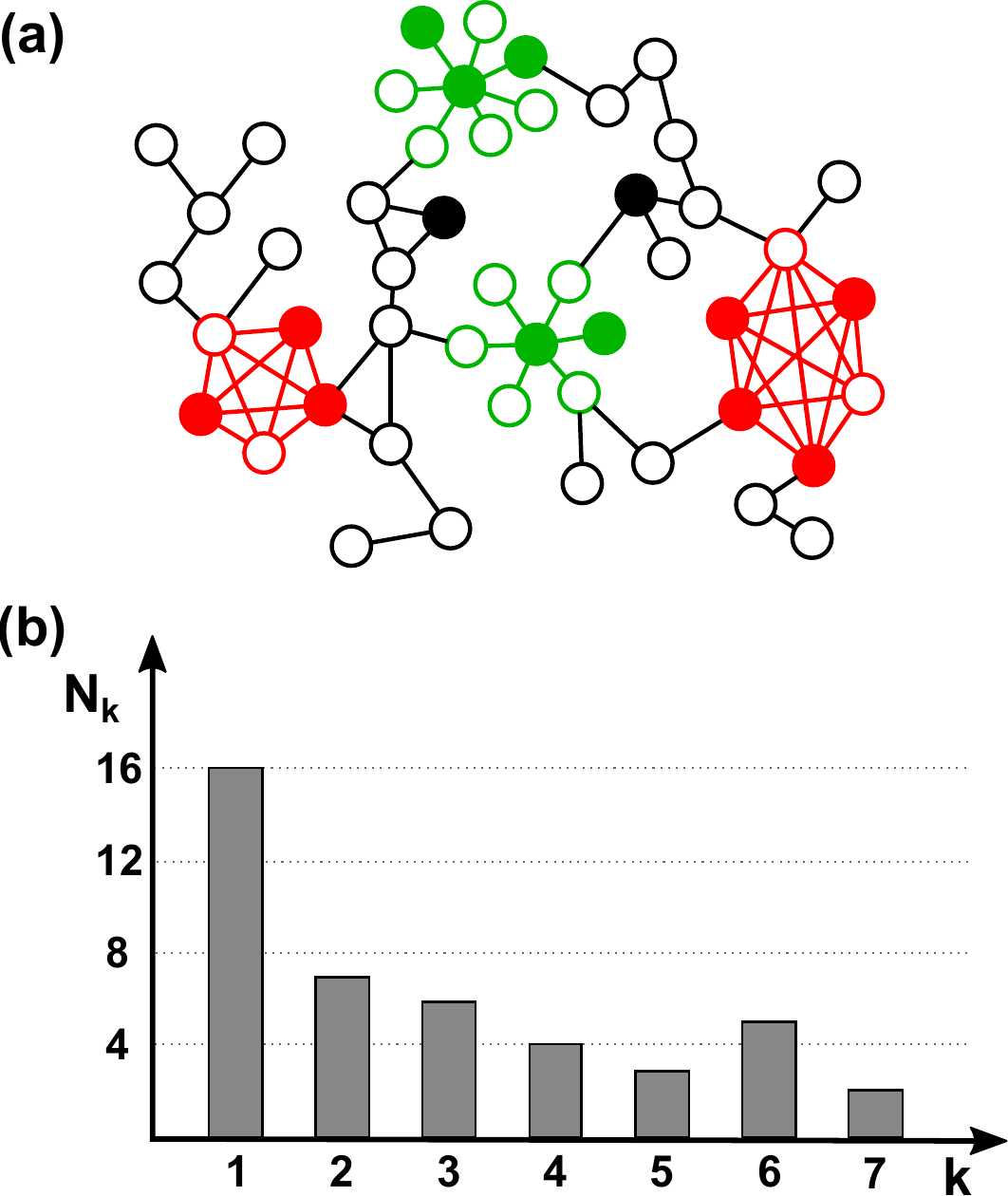}
	
	\caption{(a) Example of a network  schematically shown as  a graph, in which vertices represent individuals and edges connecting them are the possible routes for infections. (b) Degree count for the network. Two subgraphs play important roles in the epidemic activation: cliques of fully connected vertices (red) and stars consisting of hubs with its neighbors (green). Open and closed symbols represent susceptible and infected individuals in a SIS epidemic process, respectively.}
	\label{fig:scheme}
\end{figure}

Many phenomena can be suited  within the framework of dynamical processes on complex networks, in which individuals represented by nodes (or vertices)  interact with each other and the interactions are mediated by links (or edges)~\cite{Barrat2008}, as illustrated in Fig.~\ref{fig:scheme}. Nodes and links are more frequently used in the context of physical systems while vertices and edges for their mathematical representation as graphs. We use both terminologies interchangeably. Examples  range from rumor and epidemic spreading~\cite{Cota2019,Yang2019,Halu2019,Gleeson2016,Soriano2018} to urban mobility \cite{Depersin2018,Giulia2017}. In particular, those  with absorbing states, in which the dynamics is trapped forever when visited~\cite{Marro2005,Henkel2008}, represent an important class of processes on networks~\cite{Barrat2008}.  

Epidemics can be modeled as reaction-diffusion systems, in which the states of agents and the corresponding transitions are related to epidemiological stages (susceptible, infectious, immunized, etc)~\cite{Pastor-satorras2015}. One of the most fundamental epidemic models is the susceptible-infected-susceptible (SIS), in which individuals can be infectious or susceptible. The former heals spontaneously with rate $\mu$ and the latter is infected with rate $\lambda$ per infectious contact~\cite{Anderson1991}. On networks, the contagion is mediated by edges connecting infected and susceptible nodes.  Since the infection cannot arise spontaneously, a configuration with only susceptible individuals is an absorbing state.

The SIS model on power-law networks with degree distribution $P(k) \sim k^{-\gamma}$, defined as the probability that a randomly selected vertex has $k$ connections (Fig.~\ref{fig:scheme}), presents remarkable properties.  For example, the epidemic threshold, which separates an inactive (absorbing) from an active (fluctuating) phase, is asymptotically null~\cite{Durret2009,Boguna2013} that implies a nonzero prevalence (fraction of infected nodes) for any finite infection rate in an infinite network. For finite systems, the dynamics near to the transition is featured by a localized activation~\cite{Mata2015,deArruda2017,Moretti2013,Wei2020,Silva2021}. In the particular case of random power-law networks, the competition of  two basic localized structures, stars and maximum $k$-core subgraphs~\cite{Kitsak2010,Castellano2012}, plays a central role on the activity distribution. A star subgraph is composed of a vertex of degree $k\gg \av{k}$ plus its nearest-neighbors, see Fig. \ref{fig:scheme}, and can sustain a localized metastable activity of a SIS dynamics~\cite{Durret2009}, permitting mutual activation of hubs even if they are far from each other~\cite{Durret2009,Boguna2013,Ferreira2016a}. The maximum $k$-core~\cite{Kitsak2010}, which is the minimal connected subset containing only elements of degree $k\ge K$ after the removal of all vertices of degree $k<K$  and the edges connected to them, may undergo an activation before the star graphs being responsible by the onset of the endemic phase~\cite{Castellano2012}. However, more complex scenarios can emerge on real networks, in which other motifs, such as cliques of fully connected vertices (Fig.~\ref{fig:scheme}), rule the epidemic activation~\cite{Pastor-Satorras2018}.

Analyses of absorbing-state phase transitions  by means of simulations demand approaches to prevent trapping into absorbing states since, strictly, they are  the only actual steady-state for finite sizes~\cite{Marro2005}; the only ones accessible in simulations. A fundamental approach is  the quasistationary (QS) analysis~\cite{Marro2005}, in which averages  at a time $t$ are performed only over a subspace of configurations that did not visit the absorbing state until that time. Alternatively, one can  consider some perturbation of the dynamics in such a way that the absorbing state is no longer accessible. However, these perturbations must be negligible for  intensive and extensive quantities in the thermodynamic limit~\cite{Sander2016a}. There are some methods conceived within this approach~\cite{Lubeck2003,Pruessner2007,Dickman2002}.

The standard quasistationary which are hereafter generally called QS method (SQS) can be  implemented by  storing
configurations picked up at random along simulations with a certain rate (a
parameter of the method) and using them to form a new active state whenever the
dynamics is trapped into an absorbing state~\cite{deOliveira2005}. The method
has frequently been used to investigate localization of epidemic processes on
networks~\cite{Mata2015,deArruda2017,Valdano2017,Cota2016,Ferreira2016,St-Onge2017}. Despite being general, the method has some drawbacks for simulations on large networks such as the high RAM memory load to store a large number of configurations, metastability problems due to finite sampling and time averaging on critical and subcritical simulations, and parameters which do not have a systematic calibration and are model dependent. Simpler methods, such as reflecting boundary condition (RBC), in which the dynamics returns to the previously visited configuration if  it has fallen into an absorbing state, can circumvent these problems~\cite{Dickman2002}.  A systematic comparison between SQS and RBC~\cite{Sander2016a} for the SIS model on power-law networks shows that the latter is unable to capture some metastable dynamics related to localized states as  the former does, leading, for example, to different finite-size scaling (FSS) exponents at the transition.  A variant of the RBC, called hub reactivation~\cite{Sander2016a,Lee2013},  consists of reinfecting the most connected vertex of the network every time the absorbing state is visited. This method outperformed RBC on the identification of localized phases, equating to the SQS method. However, this method is biased for dynamics in which a single hub plays a major role in the localization and  calls for a generalization towards other dynamical processes and relevant network structures such as the cliques shown in Fig.~\ref{fig:scheme}.

In this paper, we investigate a QS method without tuning parameters and metastability issues, demanding reduced amount of RAM, and able to capture the same localization patterns on heterogeneous networks as the SQS does. It consists of reactivating a given number of vertices, chosen proportionally to their activity times along the whole history of the simulation, being hereafter called reactivation per activity time (RAT). We compare RAT with the SQS and RBC methods using the challenging  SIS dynamics on a variety of synthetic and real networks with high heterogeneity, including scale-free and multiplex structures~\cite{Bianconi2018}. Among other findings, we report that all methods provide the same epidemic thresholds in scale-networks with degree exponents $\gamma<3$, but RBC differs on the FSS of the epidemic prevalence at the epidemic threshold. Using dynamical susceptibility curves~\cite{Mata2015}, we report that the SQS and RAT methods capture the same localization patterns, even in the cases where RBC and SQS mismatch, as for example, in synthetic networks with degree exponent $\gamma>3$ and real networks with degree correlations and more complicated structures. Metastability problems due to insufficient time averaging were not observed for RAT in the networks where they were for SQS. Last but not least,   the RAM load in RAT is greatly reduced with respect to SQS while CPU times are slightly smaller.

The remainder of this paper is organized as follows. In Section~\ref{sec:qsmet} we describe the investigated QS methods and the algorithms to simulate the SIS model. We present simulations comparing the QS methods on synthetic and real networks  in section~\ref{sec:qssint}. The computer performance of the QS methods are presented in section~\ref{sec:compare}. Finally, we draw our remarks and prospects for potential applications of the RAT method in section~\ref{sec:prospects}.

\section{Methods}
\label{sec:qsmet}
Consider systems with $N$ elements labeled by $i=1,\ldots,N$ which be can either active or inactive. In the case of epidemic processes we have infected and susceptible individuals, but we keep the generic active and inactive terms allowing extension to other dynamical processes. We follow the approach of QS method as a perturbation of the dynamical rules that prevents the trapping into absorbing states in such a way that this disturbance is negligible  and does not alter any system's intensive or extensive properties in the  thermodynamic limit~\cite{Sander2016a}. However, it is important to stress that subextensive (increasing sub-linearly with $N$) and non-extensive (vanishing as $1/N$) quantities may depend on the QS method, including critical and localized phases investigated in the present work~\cite{Sander2016a}.

A central quantity to be computed is the QS distribution $P_\text{qs}(s)$ defined as the probability that $s$ elements are active in the QS regime. It is computed after a relaxation time $t_\text{rlx}$ during an averaging time $t_\text{av}$. Formally, it can be computed as 
\begin{equation}
	P_\text{qs}(s)=\frac{1}{t_\text{av}}\int_{t_\text{rlx}}^{t_\text{rlx}+t_\text{av}} \delta_{s,n(t)} dt,
\end{equation}
in which $n(t)$ is the number of active individuals at time $t$ and $\delta_{i,j}$
the Kronecker delta symbol. Computationally it  is implemented as 
\begin{equation}
P_\text{qs}(s)\mapsto P_\text{qs}(s) +\delta_{s,n(t)}\Delta t/t_\text{av},
\label{eq:Psincrement}
\end{equation}
in which  $\Delta t$ is the time step used in the Gillespie algorithm given in subsection~\ref{sub:oga}. Basic measures can be directly obtained from $P_\text{qs}(s)$ such as the QS moments~\cite{Dickman1998}
\begin{equation}
\av{\rho^k} = \frac{1}{N^k}\sum_{s=1}^{N}s^kP_\text{qs}(s).
\label{eq:qsmom}
\end{equation}
The order parameter is the QS epidemic prevalence given by $\av{\rho}$. Localization and critical regions can be identified using the QS dynamic susceptibility given by~\cite{Mata2015},
\begin{equation}
\chi = N \dfrac{\av{\rho^2}- \av{\rho} ^2}{\av{\rho}},
\end{equation}
which presents sharp peaks for values of $\lambda$ at a clean phase transition
but also detects localized activation involving sub-extensive components of the
networks in the form of additional
peaks~\cite{Mata2015,deArruda2017,Cota2016}.

The remainder of this section is devoted to  reviewing SQS and RBC methods and  to introduce the RAT method. An initial condition with all vertices infected are considered, averages are computed after a relaxation time $t_\text{rlx} = 10^7 \mu^{-1}$  during an averaging interval $t_\text{av} = 3\times 10^7 \mu^{-1}$. In very supercritical simulation these times are reduced by one order of magnitude. Here, it worths to comment that these two parameters are intrinsic to any QS method and obey the trivial rule, the larger the more accurate, which is not the case for the other parameters of the SQS method described next in subsection~\ref{sub:sqs}.

\subsection{Standard quasistationary method}
\label{sub:sqs}

The SQS method  consists of averaging restricted to the ensemble of active samples that did not visit the absorbing states up to the current simulation time~\cite{Marro2005}. For subcritical and critical analyses of finite-size systems, in which the dynamics often falls into the absorbing states, the SQS method provides only small intervals of stationary data, usually noisy and not allowing long-time series analysis. de~Oliveira and Dickman~\cite{deOliveira2005} proposed a simulation method to overcome these limitations which consists of jumping to a previously visited active configuration, every time the system falls into the absorbing state. The convergence of this method to the actual QS state is discussed in Ref.~\cite{Blanchet2016}.

The QS distribution, which is not known \textit{a priori},  must be computationally emulated. This can be done by saving copies of active configurations visited during the simulation and picking up randomly  one of them whenever the system visits an absorbing state. Two parameters are crucial in this algorithm: the number of saved configurations ($n_\text{conf}$) and the rate with which the list is updated with the current state ($p_\text{rep}$). The calibration of these parameters may be challenging given  the limited (finite) computational resources. If $n_\text{conf}$ is too small, the dynamics can be biased toward a subset which is not representative of the actual QS state; if it is too large, RAM allocation becomes an obstacle and the memory of the initial condition takes too long to be erased. If $p_\text{rep}$ is too large, saving configurations demands too much computational power and simulations can also be improperly biased  towards the latest configurations. On the other hand, if $p_\text{rep}$ is too small,  the convergence to  the QS regime takes too long due to the slow loss of memory of the initial conditions.


 \subsection{Reflecting boundary condition}
  \label{sub:rbc}
  
The RBC  method is much simpler than SQS: when the system falls into an absorbing state, it jumps to the immediately preceding configuration with a fixed rate usually chosen as the inverse of the time unit used. In the case of SIS it is just $\mu$. The equivalence between RBC and an external weak field, in which vertices are spontaneously  reactivated  with a very small rate, was investigated for regular lattices in Ref.~\cite{Pruessner2007} and networks in Ref.~\cite{Sander2016a}. It was shown that  RBC is equivalent to a random reactivation of a single vertex. Both variations are computationally simple but do not reproduce the QS distributions obtained with SQS  method and especially if there are strongly localized activations~\cite{Sander2016a}. Other approaches can be constructed considering targeted reactivation  of vertices as, for example,  the hub reactivation~\cite{Lee2013,Sander2016a} aforementioned.

\subsection{Reactivation per activity time}
 \label{sub:rat}

\begin{figure}[hbt]
\includegraphics[width=0.9\columnwidth]{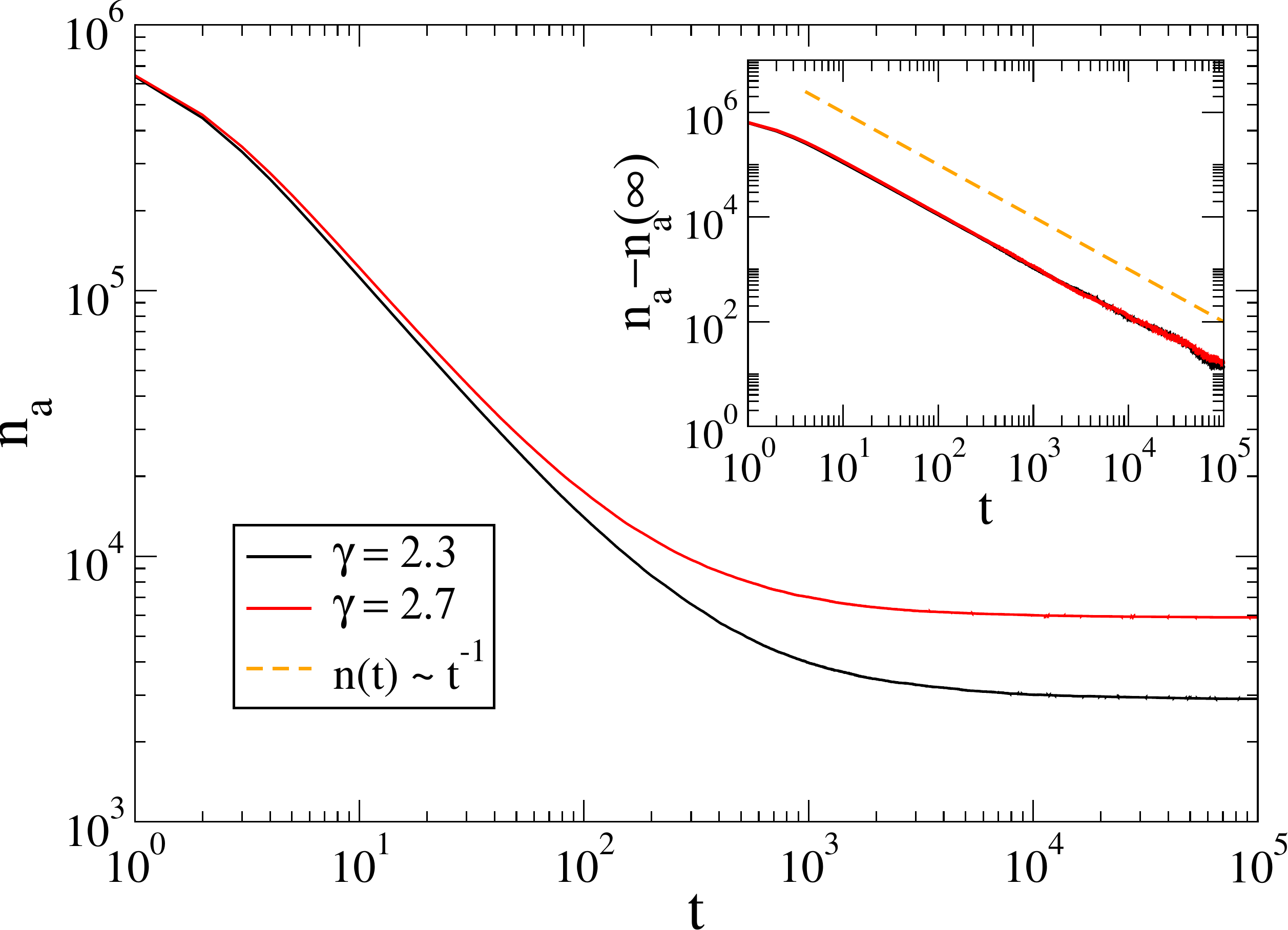}
\caption{Evolution of the average number of active individuals in SIS simulations running on power-law networks with $N=10^6$ vertices and degree exponents $\gamma=2.3$ and $2.7$. Inset shows the same data subtracted from the stationary value $n_\text{a}(\infty)$. The simulation was run for infection rates $\lambda=0.0115$ and $\lambda=0.0362$  for $\gamma=2.3$ and $\gamma=2.7$, respectively, which give the maximum value of the dynamical susceptibility; see Sec.~\ref{sec:compare} for more details.}
	\label{fig:nt}
\end{figure}

Inspired  by simple reactivation dynamics, we describe a method able to capture the activation of network subgraphs as the SQS method does. Randomly selected vertices are reactivated according to their history of activity in such a way that those which were active for longer times are more probable of being reactivated. The activity time of the vertex $i$ at time $\tau$ along a simulation is given by
\begin{equation}
T^\text{(a)}_i(t) =  \int_{0}^{t}\sigma_i(\tau)d\tau,
\label{eq:Ta}
\end{equation}
in which $\sigma_i=1$ if vertex $i$ is active and $\sigma_i=0$, otherwise.
After the  dynamics has fallen into an absorbing state,  $\lceil n_\text{a}\rceil$ or  $\lfloor n_\text{a} \rfloor$ vertices are reactivated with probabilities $ n_\text{a}-\lfloor n_\text{a} \rfloor$ and $\lceil n_\text{a} \rceil-n_\text{a}$, respectively, in which $n_\text{a}$ is the overall average activity computed as
\begin{equation}
n_\text{a}(t) = \frac{1}{t}\sum_{i=1}^{N}T^\text{(a)}_i
\label{eq:na}
\end{equation}
while $\lfloor x \rfloor$  and $\lceil x \rceil$ are the floor and ceiling functions, respectively. So, the relaxation time must be large enough to erase the influence of the initial condition. For all simulations performed in the present work the convergence is very fast as exemplified for the SIS dynamics near the transition in Fig.~\ref{fig:nt}.

Vertices to be reactivated are chosen at random proportionally to their total activity time. This selection can be done using the rejection method, in which a randomly chosen vertex is accepted with probability
\begin{equation}
q_i= \frac{ T^\text{(a)}_i }{ \max\limits_{j} \left\{ T^\text{(a)}_j \right \} }, 
\end{equation} 
which may be computationally costly if the system falls very often into an absorbing state and the activation time $T^\text{(a)}_{i}$ has a broad distribution with values of $T^\text{(a)}_i\gg \av{T^\text{(a)}_i}$. This difficulty is circumvented  by dividing the vertices in $Q$ groups $\mathcal{G}_g$, $g=1,\ldots,Q$  according to some property related to its activity.  Next, the total activity time of each group given  by
\begin{equation}
T_{g} = \sum_{j\in \mathcal{G}_g} \left(T^\text{(a)}_j \right),
\end{equation} 
is computed and  one of the groups selected with probability
\begin{equation}
P_g=\frac{T_{g}}{\sum\limits_{g^\prime=1}^Q T_{g^\prime}}.
\end{equation}
Finally, a randomly selected vertex $i\in \mathcal{G}_g$  is accepted with probability 
\begin{equation}
q_i=\dfrac{T^\text{a}_i}{\max\limits_{j\in  \mathcal{G}_g}\lbrace T^\text{a}_j\rbrace}.
\end{equation}
The number of groups and criteria to divide them should be adapted to the dynamical
process and network under investigation. 

The vertex degree is a natural choice for the  SIS model or other processes
where activation is related with vertex connectivity. So, in the present work, vertices were labeled such that $k_1\le k_2\le,\cdots,\le k_N$ to form groups of equal size $M=N/Q$ given by
\begin{equation}
\mathcal{G}_g=\lbrace k_i| (g-1)M+1\le i\le g M \rbrace,~g=1,\cdots,Q.
\label{eq:Gg}
\end{equation}
In the case of multiplex networks (section~\ref{sub:multi}), the binning can be done separately for each layer. This approach will reduce the number of rejections by gathering vertices of similar degree in the same bin. Except for the networks of section~\ref{sub:rrnhub}, which has a single outlier considered separately, $Q=100$ was adopted for all simulations in this paper.

\subsection{Algorithms for SIS model}
\label{sub:oga}

To simulate the SIS dynamics on simple connected networks, we used the optimized Gillespie algorithm (OGA) with phantom processes described in Ref.~\cite{Cota2017}; See also~Ref.~\cite{St-Onge2019} for similar methods. The number of infected vertices $\Ninf$ and total number of edges connected to them $\Ne$ given formally by
\begin{equation}
\Ninf = \sum\limits_{i=1}^{N} \sigma_i ~\text{and}~ \Ne=\sum\limits_{i=1}^{N} \sigma_i k_i,
\end{equation}
are determined and kept updated along the simulations. With probability 
\begin{equation}
p = \frac{\mu\Ninf}{\mu\Ninf + \lambda \Ne},
\end{equation} 
an infected vertex chosen at random becomes susceptible. With complementary probability $w=1-p$ an infected vertex is  selected at random, proportionally to its degree, and one of its neighbors is selected at random with equal chance. If the selected link points to a susceptible vertex, it  is infected. Otherwise, the simulation goes to the next step without altering the configuration (phantom process). Finally, the time is incremented by 
\begin{equation}
\Delta t = \frac{-\ln \xi}{\mu \Ninf + \lambda \Ne},
\label{eq:Delta_t}
\end{equation} 
where $\xi$ is uniformly distributed in the interval $(0,1]$. This whole process is iterated for a predefined time. The activity times of each vertex can be exactly computed by storing the last time it became infected time which is subtracted from the current  healing time. However, since the healing is spontaneous, this step can be simplified by incrementing $T_i^\text{(a)}$ by $1/\mu$ every time the vertex $i$ is healed.

We also analyzed SIS models on multilayer networks,  in which same agent belongs to different networks (layers) whose edges represent distinct interactions~\cite{Bianconi2018}. For example, a same individual who has Twitter and Facebook profiles with different sets of followers. Let us consider a  network with $\alpha=1,2,\cdots,m$ layers; each layer is connected to all others. SIS processes with the same parameters $\mu$ and $\lambda$ evolve in each layer. Additionally, a vertical transmission rate $\eta$, in which one vertex in a layer activates itself in another layer~\cite{deArruda2017}, is included.  The OGA for the SIS dynamics on multilayer networks runs as follows. With probability
\begin{equation}
p=\dfrac{\mu\Ninf}{(m-1)\eta \Ninf + \mu\Ninf + \lambda \Ne},
\label{eq:phealMplx}
\end{equation}
an infected vertex selected at random is healed. With probability 
\begin{equation}
q = \dfrac{(m-1)\eta \Ninf}{(m-1)\eta \Ninf + \mu\Ninf + \lambda \Ne},
\end{equation}
an infected vertex is randomly chosen and activates itself in another layer, also chosen at random, if it was not active yet. Finally, with probability $w=1-q-p$ we perform the intralayer infection process of the SIS model described for a single layer. The time increment $\Delta t$ is analogous to Eq.~\eqref{eq:Delta_t} with the proper denominator given in Eq.~\eqref{eq:phealMplx}. Note that a vertex active in multiple layers have to be considered accordingly such that
\begin{equation}
\Ninf = \sum_{\alpha=1}^{m}\sum_{i=1}^{N}\sigma_{i\alpha} ~\text{and}~\Ne=\sum_{\alpha=1}^{m}\sum_{i=1}^{N}k_{i\alpha}\sigma_{i\alpha},
\end{equation}
in which $\sigma_{i\alpha}$ and $k_{i\alpha}$ indicate, respectively, the states and degree of vertex $i$ in the layer $\alpha$.

\section{Comparison of QS methods}
\label{sec:qssint}

We performed QS simulations of the  SIS model on complex networks with broad levels of localization to compare the different methods. For simplicity we adopt $\mu=1$, fixing therefore the time unit without loss of generality. We start with simple graphs where the localization is highly controlled in subsection~\ref{sub:rrnhub}. Then, we consider increasing complexity with synthetic power-law, multiplex, and real networks in subsections~\ref{sub:pl}, \ref{sub:multi}, and \ref{sub:real}, respectively.

\subsection{Random regular networks with a hub}
\label{sub:rrnhub}

\begin{figure}[hbt]
	\includegraphics[width=0.9\linewidth]{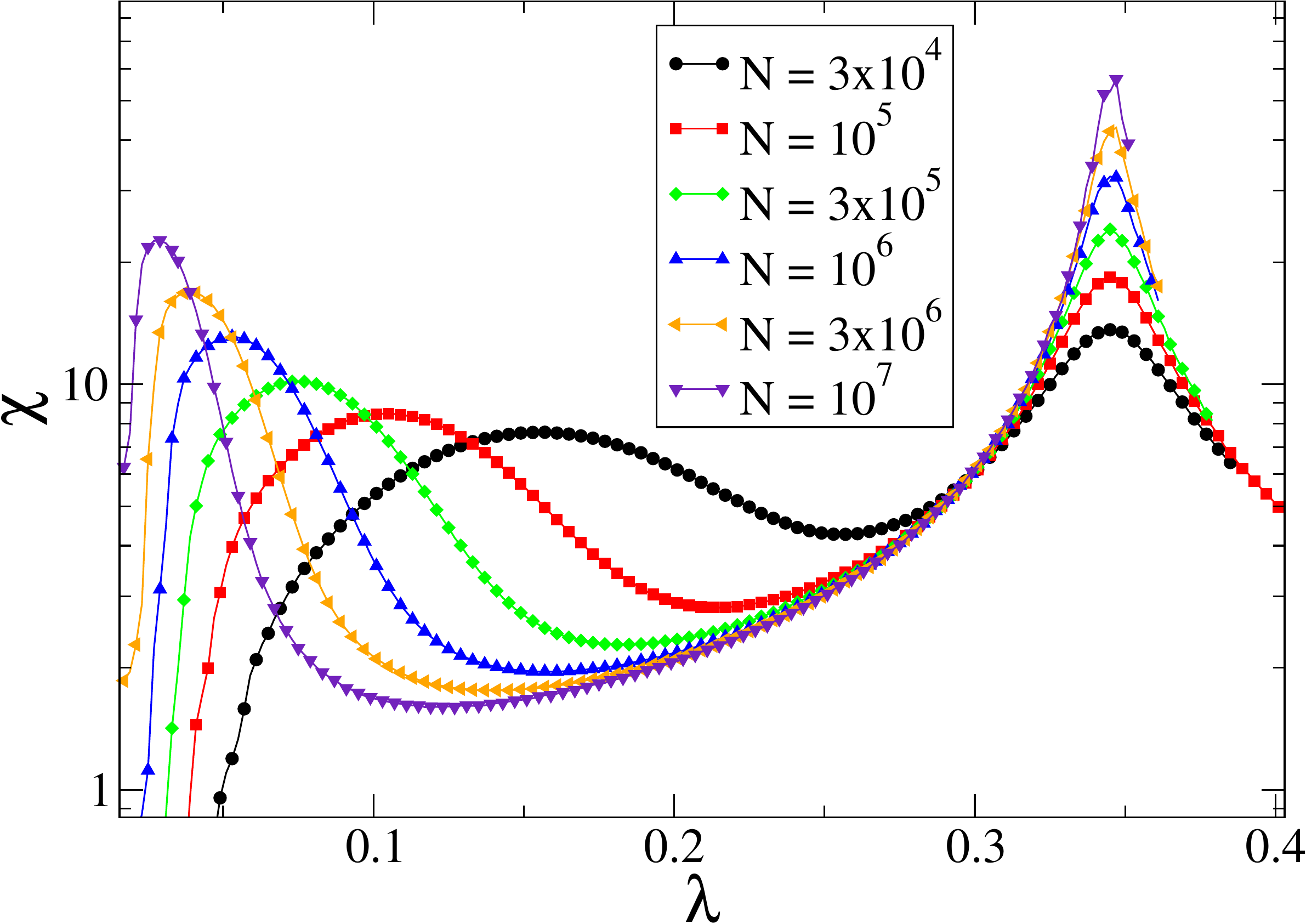}
	\caption{Susceptibility as a function of the infection rate for RRN ($m=4$), of different sizes $N$ (indicated in the legends) plus a hub of degree $K=\sqrt{N}$ using the RAT method.}
	\label{fig:rn1h-rta}
\end{figure}
\begin{figure}[hbt]
	\includegraphics[width=0.95\linewidth]{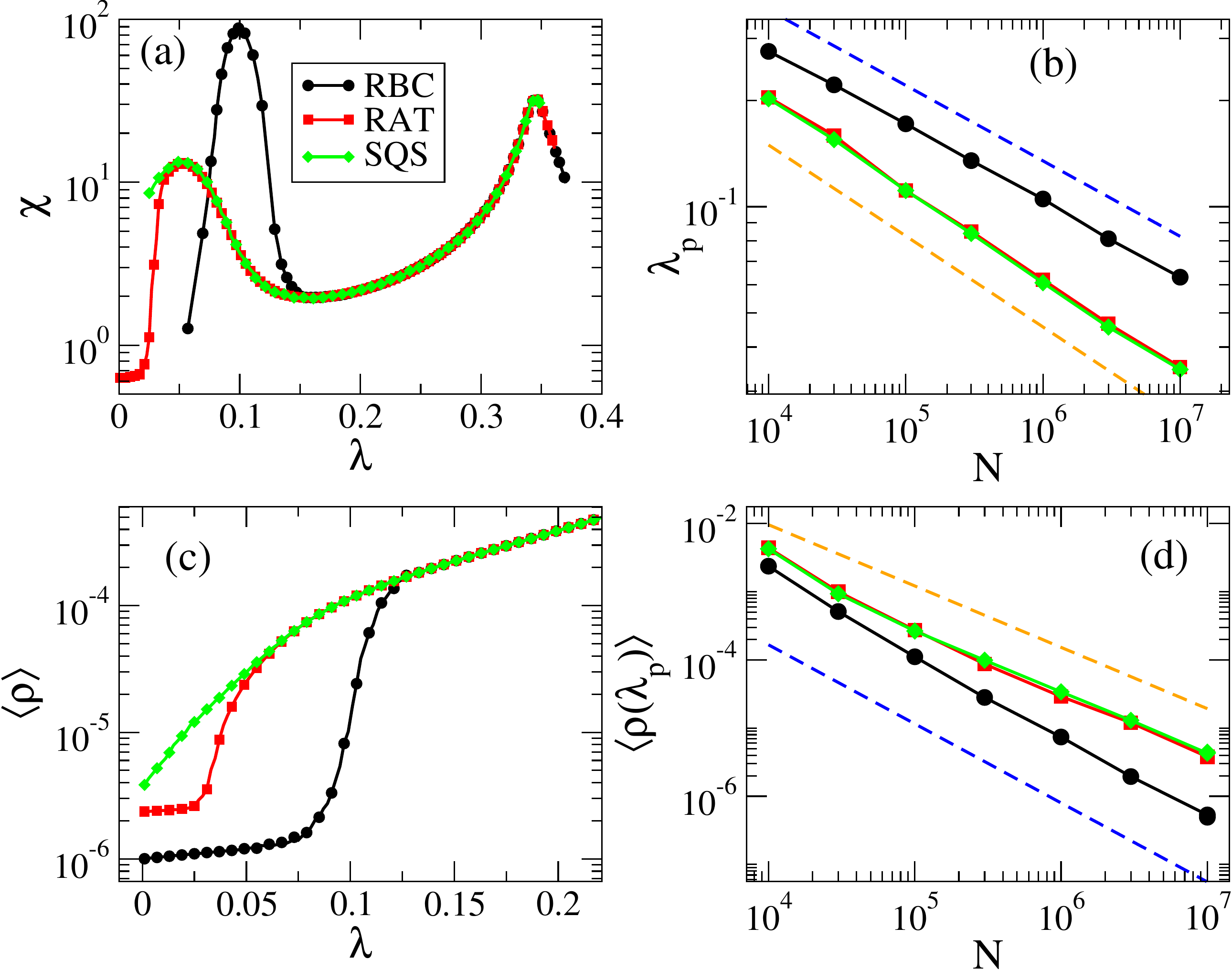}
	\caption{Quasistationary quantities for a RRN ($m=4$) with $N= 10^6$  vertices plus a hub of degree 	$K=10^3$ obtained with  different QS methods. (a) Susceptibility and (c) prevalence versus infection rate.  Finite-size scaling for the (b) position of left susceptibility peak and (d) the corresponding QS prevalences. Dashed lines represent power-laws with exponents $-0.27$ 	(top) and $-0.21$ (bottom) in (b) and $-0.90$ (top) and $-1.2$ (bottom) in (d). }
	\label{fig:rn1hcomp}
\end{figure}

Random regular networks (RRN) are simple graphs, in which every vertex has the same degree ($m$) and the connections are performed at random  without multiple or self-connections. The SIS model on RRNs undergoes a phase transition at a finite epidemic threshold~\cite{Ferreira2013}. Introducing a single vertex of degree $K\gg m$ in a RRN and still connecting edges at random, a localized activation at $\lambda\sim 1/\sqrt{K}$, in which only the hub and its neighborhood  are involved, is expected in addition to the global one at a finite $\lambda$~\cite{Ferreira2016}. In Fig.~\ref{fig:rn1h-rta}, one can see both activations  in the form of peaks in  the curves of  susceptibility versus  infection rate for the SIS model on RRNs of degree $m=4$ plus one hub of degree $K = \sqrt{N}$ for different network sizes using the RAT method. Peaks moving towards zero represent the localized activation, while those accumulating at a finite value represent the global ones.

Figure~\ref{fig:rn1hcomp} compares some quantities of interest obtained with the three QS methods for SIS on a RRN of $N=10^6$ and a hub of degree $K=10^3$.  The susceptibility and prevalence curves, shown in Fig.~\ref{fig:rn1hcomp}(a) and (c), obtained with RAT and SQS  match almost perfectly each other except in the very subcritical region where $\av{\rho}\sim 1/N$. In contrast, the localized activation of the hub, characterized by the left peak of the susceptibility curve at $\lambda=\lambda_\text{p}$, identified with the RBC method is distorted with respect to the SQS method. A finite-size scaling (FSS) of the position $\lambda_\text{p}$  of the left peak in the susceptibility curves, which measures the localized activation of the  hub, is presented in Fig.~\ref{fig:rn1hcomp}(b). While RAT and SQS present the same values for $\lambda_\text{p}$ within uncertainties, RBC deviates substantially. Assuming a scaling law of the form $\lambda_\text{p} \sim N^{-\phi}$, a same exponent  $\phi = 0.21$ is obtained for both RAT and SQS  whereas $\phi = 0.26$ for RB. The same FSS analysis can be done for the QS prevalence at $\lambda=\lambda_\text{p}$,  Fig.~\ref{fig:rn1hcomp}(d), and a similar result is found, in which the scaling obtained with RBC, $\langle \rho(\lambda_\text{p})\rangle \sim N ^{-1.20}$, differs from those of SQS and RAT, $\langle  \rho(\lambda_\text{p})\rangle \sim N ^{-0.90}$.

\subsection{Synthetic power-law networks}
\label{sub:pl}
Many real networks have a heavy-tailed degree distribution following a power-law $P(k) \sim k^{-\gamma}$~\cite{barabasi2016network}. We investigate the performance of the QS methods on power-law, synthetic networks constructed using the uncorrelated configuration model (UCM)~\cite{Catanzaro2005}, in which an upper structural cutoff in the degree distribution $k_\text{c}=\sqrt{N}$ is used to guarantee the absence of degree correlations~\cite{Boguna2004}. The minimal degree $k_\text{min}=3$ was fixed in all simulations. Values of $\gamma$ related to the different activation mechanisms of the SIS model were considered~\cite{Castellano2012}: $\gamma=2.3$ for which activation is governed by the maximum $k$-core; $\gamma=2.8$ where hubs rule the activation but due to the structural cutoff, the degree distribution does not have outliers that generate multiple peaks in the susceptibility curves~\cite{Mata2015}; and $\gamma=3.5$ for which activation is also ruled by hubs, but strong localization in a few set of hubs is observed~\cite{Mata2015,Cota2016}.

\begin{figure}[hbt]
	\includegraphics[width=0.9\linewidth]{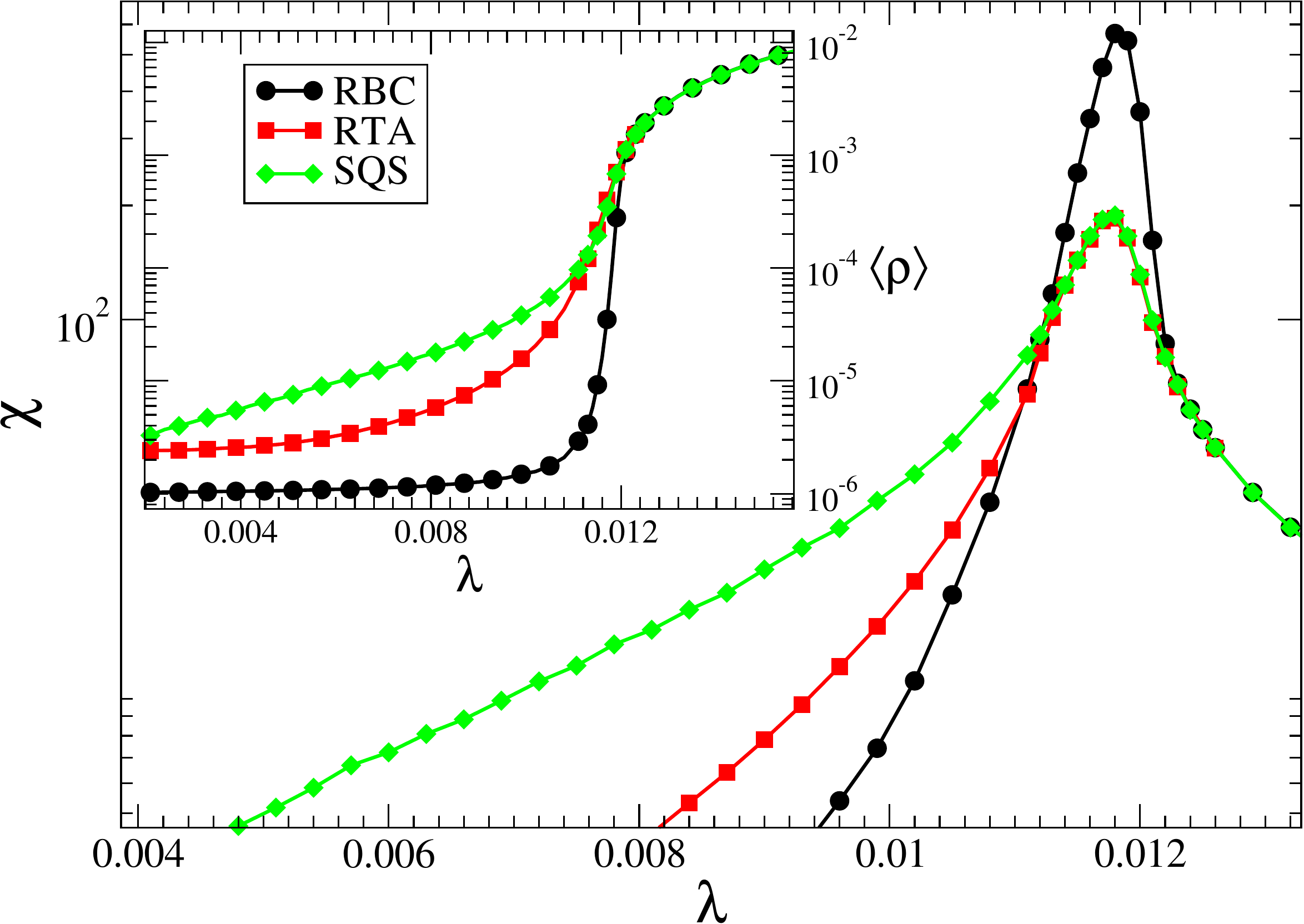}
	\caption{Susceptibility versus infection rate for a UCM network with $N = 10^6$ vertices and degree exponent $\gamma = 2.3$, considering three QS methods. Inset shows the QS prevalence as a function of the infection rate.}
	\label{fig:SUSucm23}
\end{figure}

\begin{figure}[h]
	\includegraphics[width=0.98\linewidth]{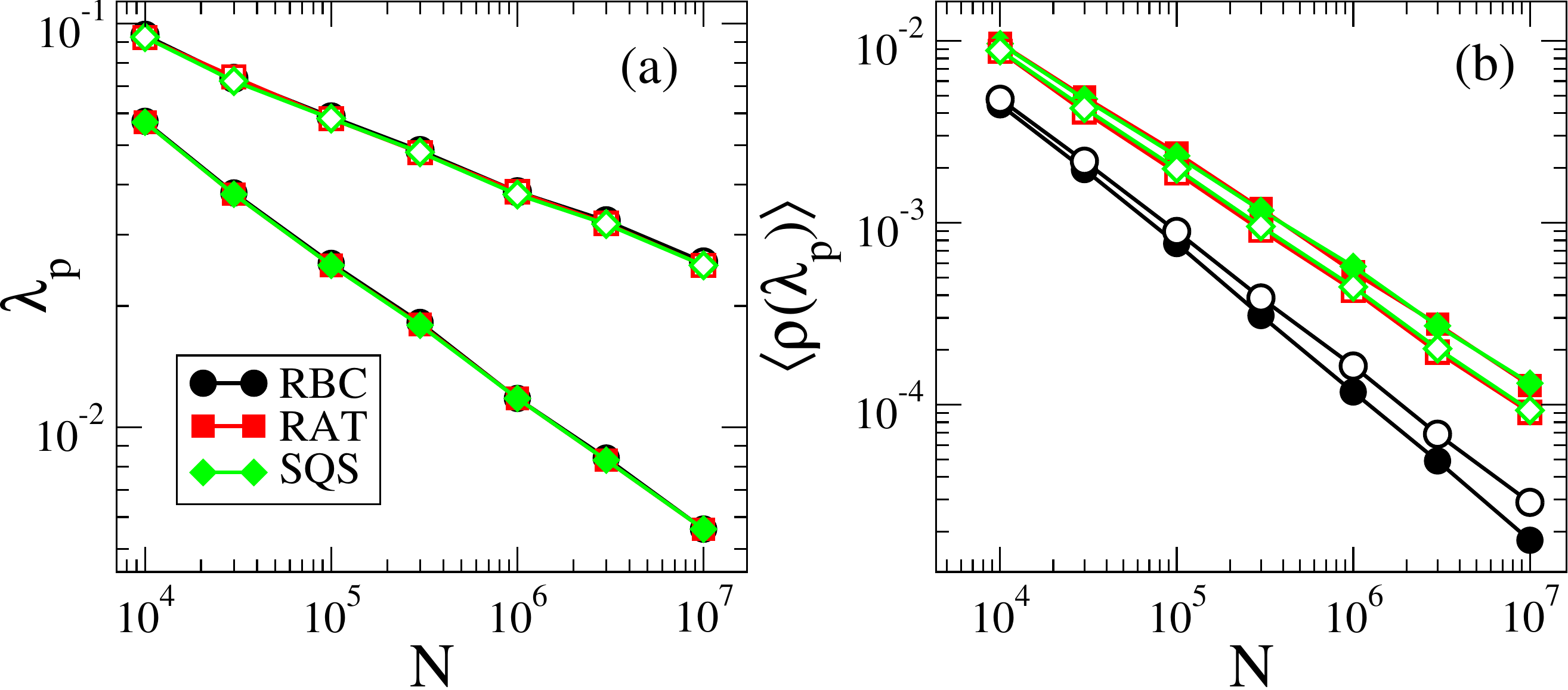}
	\caption{Finite-size scaling for the (a) epidemic threshold and (b) the corresponding QS prevalence of the SIS model on UCM networks with $\gamma = 2.3$ (closed symbols) and $\gamma = 2.8$ (open symbols).  Averages were computed over 35 networks and error bars are smaller than symbols.	}
	\label{fig:ucm28}
\end{figure}

Figure~\ref{fig:SUSucm23} presents the susceptibility and QS epidemic prevalence versus infection rate computed using different SQS methods on power-law networks with  $N = 10^6$ vertices and  degree exponent $\gamma = 2.3$. The corresponding curves for $\gamma = 2.8$ are qualitatively similar and not shown for sake of brevity. Curves for the RAT method deviate from the SQS ones only for the very subcritical phase while the RBC deviates on both the subcritical and critical phases. All methods are equivalent in the very supercritical region since absorbing states are not visited for the investigated simulation times. Figure~\ref{fig:ucm28} presents the FSS of the epidemic thresholds and QS prevalence, considering two values $\gamma=2.3$ and 2.8. All methods provide equivalent epidemic thresholds with the same scaling for both $\gamma$ values, while  the scaling of the prevalence using RBC deviates from RAT and SQS method; the last two match each other almost perfectly.

\begin{figure}[h]
	\includegraphics[width=0.98\linewidth]{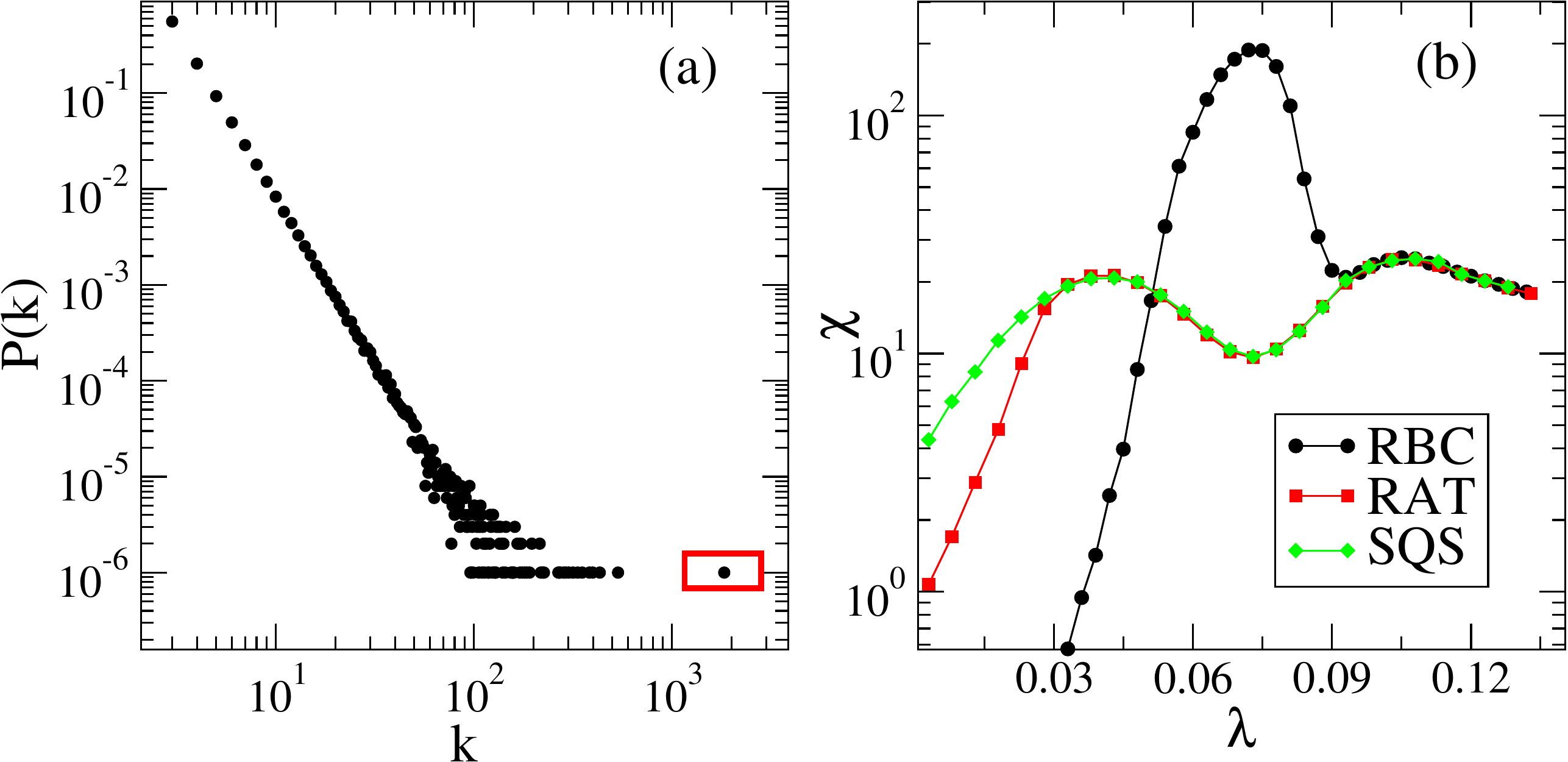}
	\caption{(a) Degree distribution of a network with $\gamma =3.5$ possessing an outlier of degree $1837$ highlighted in the box. (b) Susceptibility curves obtained using different QS methods for the network whose degree distribution is shown in (a).}
	\label{fig:ucm35}
\end{figure}

To tackle stronger localization effects, simulations on a synthetic network with $N = 10^6$  vertices and degree exponent $\gamma =3.5$ was considered. We selected a sample with an outlier of degree $1837$, as can be seen in the degree distribution in Fig.~\ref{fig:ucm35}(a). Comparing curves obtained with different methods in Fig.~\ref{fig:ucm35}(b), similarly to the case of a RRN plus a hub, RAT and SQS methods are equivalent except in a very subcritical  region while RBC deviates substantially in the whole interval of $\lambda$ where epidemic activity is highly concentrated in the outlier's neighborhood, including the position of the peak which is significantly displaced. All  methods match in the secondary peak that corresponds to the global activation of the epidemics since falling into the absorbing state is extremely rare, due to activation of the hub.

\subsection{Multiplex networks}
\label{sub:multi}
 
\begin{figure}[hbt]
	\centering
	\includegraphics[width=0.8\linewidth]{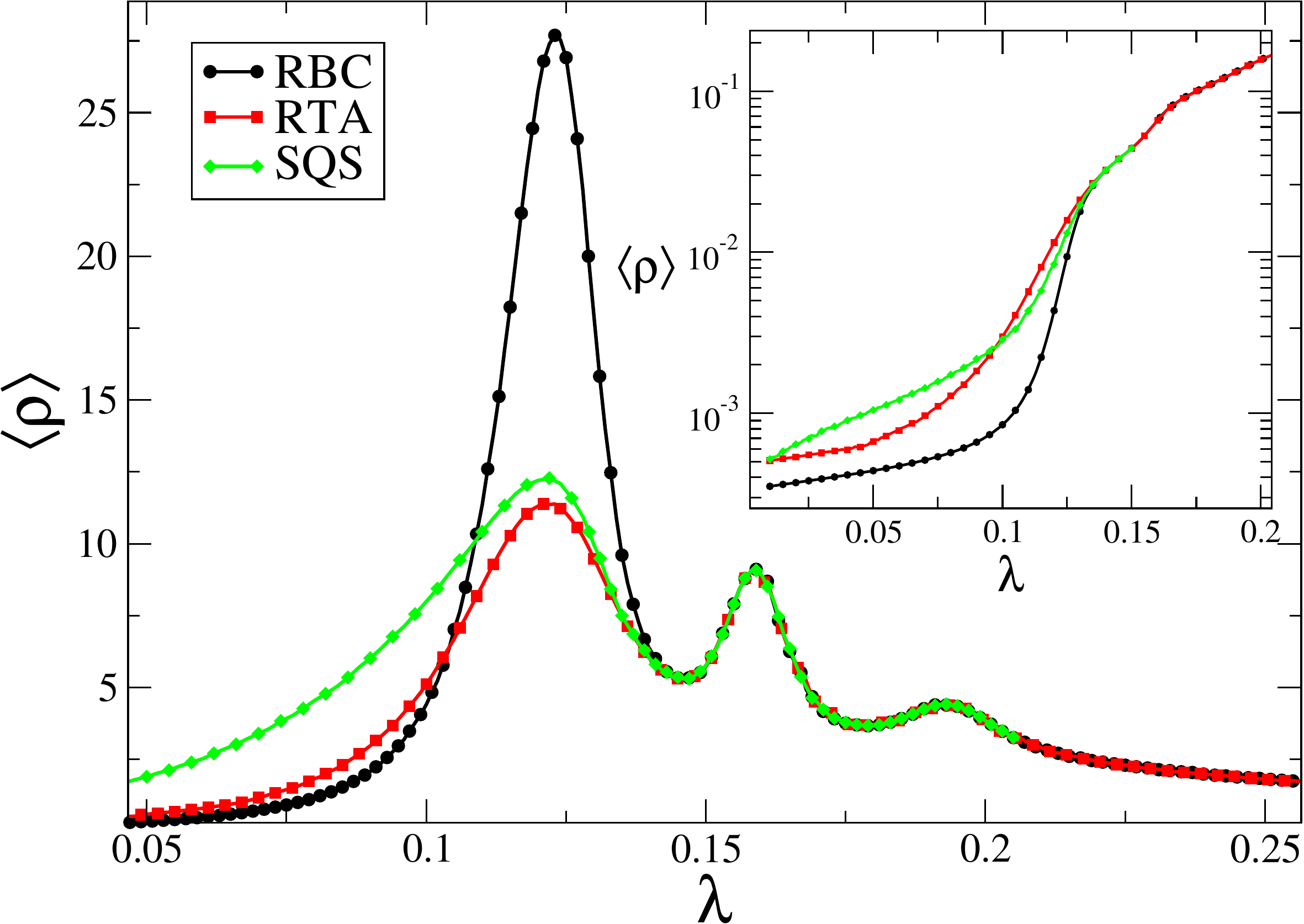}
	\caption{Susceptibility as function of the infection rate for a multiplex network with three layers ($\gamma_1 = 2.3$, $\gamma_2 = 2.6$, and $\gamma_3 = 2.9$), $\frac{\eta}{\lambda}=0.002$ and $N = 10^4$ considering three QS methods. Inset: prevalence as function of the infection rate.}
	\label{fig:multi}
\end{figure}

Simulations of the SIS model on multiplex networks show that multiple transitions related to the activation of each layer individually take place for a sufficiently small ratio $\frac{\eta}{\lambda}$~\cite{deArruda2017}. So, multiplex networks are perfect substrates to investigate localization. Figure~\ref{fig:multi} compares the susceptibility curves for a multiplex network with three layers with power-law degree  distributions of exponents $\gamma = 2.3$, $2.6$, and $2.9$. Each layer has $N = 10^4$ vertices while the ratio $\frac{\eta}{\lambda} = 0.002$ and minimal degree $k_\text{min}=3$ were fixed, such that the layer with smaller degree exponent is activated first. All QS methods capture the transitions in each layer, approximately at the same position. Small differences between SQS and RAT curves are observed before the activation of the first layer ($\gamma=2.3$) whereas RBC deviates strongly also around the first susceptibility peak.

\subsection{Real networks}

\label{sub:real}
We performed simulations of the SIS model on some real networks that, differently from UCM, present degree correlations. Six networks were chosen from  Ref.~\cite{Radicchi2015} based on exotic behaviors of the susceptibility curves generated with the SQS method. Some important properties of the networks are shown in Table~\ref{tab:propreal}. The Pearson coefficient defined as~\cite{newman2010networks}
\begin{equation}
P = \frac{\sum_{ij}\left(A_{ij}-\frac{k_ik_j}{N\av{k}}\right)k_ik_j}
{\sum_{ij}\left(k_i\delta_{ij}-\frac{k_ik_j}{N\av{k}}\right)k_ik_j},
\end{equation}
in which the adjacency $A_{ij}=1$ if vertices $i$ and $j$ are connected and  $A_{ij}=0$ otherwise, quantify the overall network topological correlations. It lays within the interval $-1\le P \le 1$, in which positive, null and negative values are related to assortative, neutral and disassortative mixing, respectively~\cite{newman2010networks}. The modularity coefficient $Q$ lays  in  the range  $[0,1]$ and quantifies  the structure of communities present in the network~\cite{newman2010networks,barabasi2016network}. The higher  modularity coefficient the more significant are  the connections among individuals of a same community. The selected networks present a wide range of sizes, heterogeneities (quantified by $\av{k^2}/\av{k}$~\cite{barabasi2016network}), degree correlations (related to $P$) and are highly modular. Highly modular structures with connectivities within  modules differing from each other are expected to present independent activation of the communities and multiple peaks in susceptibility curves~\cite{Mata2015,Cota2018b}.

\begin{figure*}[hbt]
	\centering
	\includegraphics[width=0.60\linewidth]{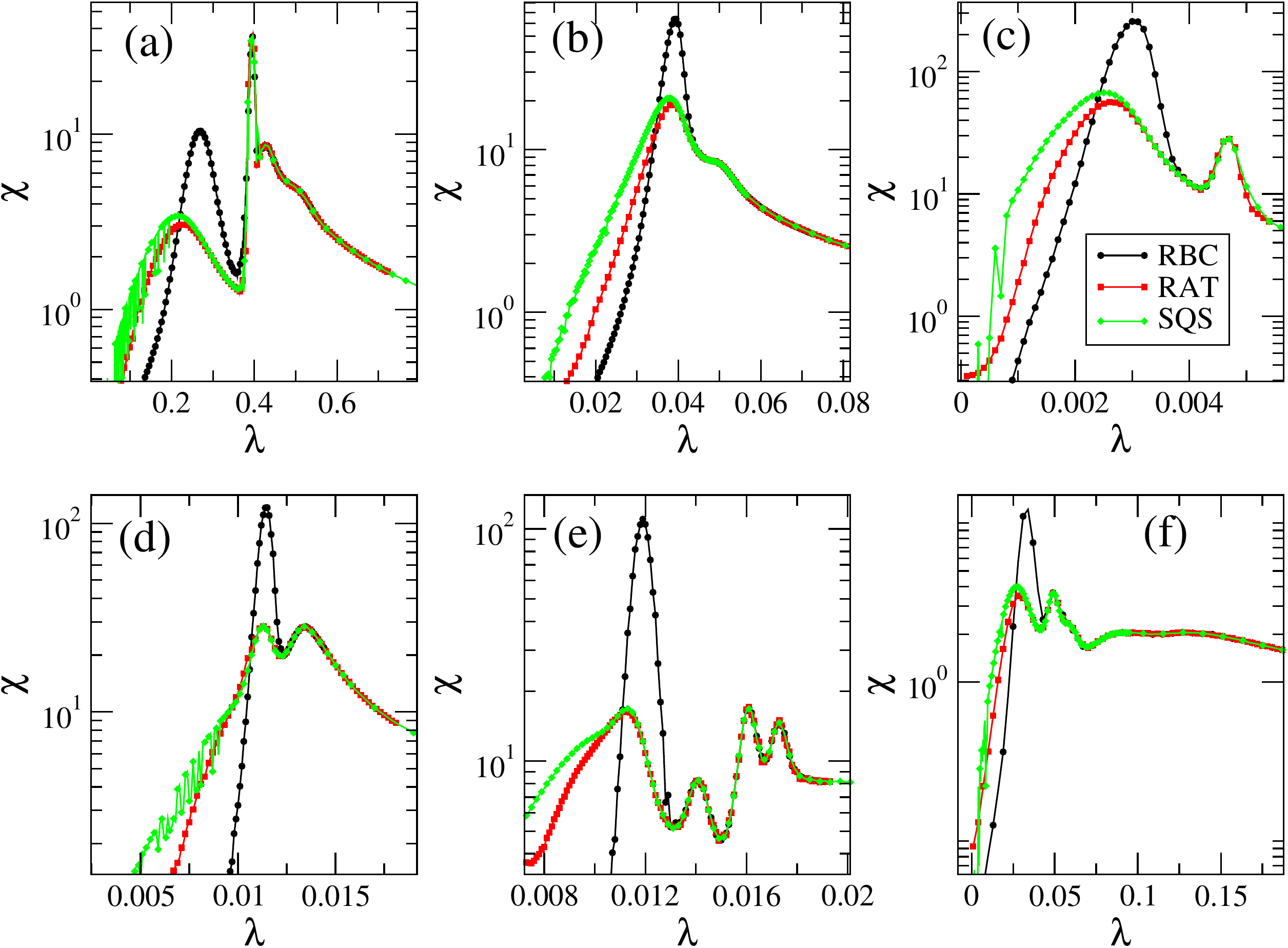}
	\caption{Comparison of susceptibility curves obtained for the SIS model on different real networks using different QS methods. Results for the networks given in table~\ref{tab:propreal} are presented: (a) USPowerGrid; (b) Cora, (c) WebStanford, (d) DBLP-Colab, (e) USPatents and  (f)		GR-QC.}
	\label{fig:realnets}
\end{figure*}
\begin{table}[hbt]
	\begin{tabular}{ccccccc}
		\hline
		Network            &   N     & $\av{k}$ & $\av{k^2}$ & $P$    & $Q$        \\
		\hline
		USPowerGrid        & 4941    & 2.66     & 10.33      & 0.003  & 0.935	   \\
		GR-QC-1993-2003    & 4158    & 6.46     & 116.1      & 0.639  & 0.848      \\
		Cora	           & 23166   & 7.69     & 182.3      & -0.055 & 0.790      \\
		DBLP-Colab         & 1137114 & 8.82     & 366.6      & 0.096  & 0.751      \\
		USPatents          & 3764117 & 8.77     & 187.8      & 0.167  & 0.812      \\
		WebStanford        & 255265  & 15.21    & 30897      & -0.115 & 0.920      \\  
		\hline		
	\end{tabular}
	\caption{Structural properties of real networks. From left to right: name of the network, number of vertices, average degree, second moment, Pearson coefficient, and modularity.}
	\label{tab:propreal}
\end{table}

As observed in  Fig.~\ref{fig:realnets}, all QS methods are able to capture the same number peaks and approximately at the same location. However, as reported for synthetic networks,  the RBC distorts the first susceptibility peak whereas RAT matches the SQS curves.

\section{Comparison on the computer efficiencies of QS methods}
\label{sec:compare}

We now discuss the computational complexity of the QS  methods. The RAT computational complexity is expected to lie between the other methods.  We analyze two key ingredients of computational complexity: RAM load and CPU time. Simulations of the SIS model on UCM networks for $\lambda=\lambda_\text{p}$ (the position of the maximum value of the susceptibility curves) were run on a workstation with an Intel i5 8400 - 2.80 GHz processor and 8 GB of RAM memory. Besides its physical relevance, the choice of the transition  point is because simulations below it are not computationally demanding while above it the QS methods become expendable since the absorbing state is almost never visited. Codes were written in Fortran~90 and compiled using a non-commercial version of the Intel Fortran for Linux. Codes for QS simulations of SIS using  RAT were made available \cite{codes} in both Fortran~90 and Python.

We evaluated the amount of RAM necessary to simulate the SIS model on a UCM network of degree exponent $\gamma=2.3$ and size $N = 10^7$. For the SQS simulations with $n_\text{conf}=100$ stored copies, up to $5$~GB of RAM memory were necessary  while for RAT and RBC  methods it did not exceed $1$ GB. This difference is due to the $n_\text{conf}\times N$ array to store several copies of the system configurations in the SQS method which are replaced by a simple array with $N$ elements to store the activity times in RAT. In SIS simulations, one could use an array smaller than $n_\text{conf}\times N$ since only the labels of the infected vertices have to be recorded to restart the simulation with a stored configuration. However, we performed this test considering the full array which is necessary for other dynamical processes as, for example, those with infinitely many absorbing states~\cite{Goh2003,Sander2013} or other epidemic processes~\cite{Ferreira2016a}.

Comparing CPU times of the SIS dynamics running for a total time $t_\text{rlx}+t_\text{av}$ misleadingly favors RBC, that takes a considerably shorter CPU time than RAT and SQS, which, in turn, are similar. Note, however, that the number of active vertices of the RBC near to the activation transition ($\lambda=\lambda_\text{p}$) is much smaller than in RAT and SQS which are essentially the same; see Fig.~\ref{fig:ucm28}. Since the time step is larger for a lower number of active vertices, Eq.~\eqref{eq:Delta_t}, less operations by unity of time are executed.  A side effect is that much less configurations are visited by unit of time and simulations using RBC are noisier than RAT for the same averaging time. For a fair comparison, we performed QS simulations of SIS at $\lambda=\lambda_\text{p}$ using UCM networks of different sizes and considering the same relaxation time $t_\text{rlx}=10^7~\mu^{-1}$. Then, we computed the averages over $10^9$ more time steps (not time units), remarking that the size $\Delta t$ is not the same for the different QS methods. In other words, the same statistics requires different averaging times for different methods. The CPU times presented in table~\ref{tab:compute} shows that all methods possess similar efficiencies where RBC is the most efficient  while RAT outperforms SQS.

\begin{table}[hbt]
\begin{tabular}{cccc}
	    \hline
	    Size       &  ~ RBC ~ & ~ RAT ~ &  ~ SQS ~ \\\hline
		$N = 10^4$ & $5.46$  & $6.63$  & $7.04$     \\
		$N = 10^5$ & $7.18$  & $8.25$  & $8.79 $     \\
		$N = 10^6$ & $10.26$ & $11.54$ & $12.44 $   \\
		$N = 10^7$ & $16.73$ & $18.19$ & $19.21 $  \\ \hline
	
\end{tabular}
\caption{CPU times (in minutes) for the QS simulations of critical ($\lambda =\lambda_\text{p}$) SIS model on UCM networks with degree exponent $\gamma=2.3$. The relaxation time is $t_\text{rlx}  = 10^7 \mu^{-1}$ and the averages were computed over $10^9$ time steps.}
\label{tab:compute}
\end{table}

The SQS method has two parameters ($p_\text{r}$ and $n_\text{conf}$) more than RBC and RAT, which are used for saving and replacing copies of previously visited configurations. A poor sampling  may  trap  the dynamic onto metastable states leading to spurious behaviors, particularly common in, but not restricted to, subcritical simulations. Such an effect can be seen in Figs.~\ref{fig:realnets}(a), (c) and (d) where non-smooth susceptibility curves are observed before the first activation transition. Finally, the code implementation for RAT is algorithmically much less complex than SQS method since the update of the list with previous configurations is replaced by a simple operation $T^{(a)}_i \mapsto T^{(a)}_i+1/\mu$ when  vertex $i$ is inactivated.

\section{Conclusions and prospects}
\label{sec:prospects}

Dynamical processes with absorbing configurations, which once visited trap the dynamics indefinitely~\cite{Marro2005}, are relevant for many biological, chemical and physical systems. In particular, the study of these processes on non-regular  structures, such as complex networks~\cite{Barrat2008}, constitutes an interdisciplinary issue with  applications to real cases~\cite{Depersin2018,Zhang2017,Merler2016,Costa2020}. Considering spreading on networks~\cite{Pastor-satorras2015}, computational simulation is  a central actor in the validation of approximated theories and in the setting up of new insights on open discussions and interpretations \cite{St-Onge2017,Boguna2013,Silva2019,Pastor-Satorras2018,Ferreira2016a}.

Simulations of active steady-states  for dynamics with absorbing configurations on finite-size systems and its extrapolation to the thermodynamic limit constitute a challenge  since  the system always visits an absorbing state for sufficiently long times~\cite{Marro2005}.  To circumvent this difficulty, one relies on the QS analysis consisting of a perturbation of the dynamical rules, negligible in the  thermodynamic limit, which prevents the system from getting trapped into the absorbing states. There are different methods to avoid the absorbing configurations such as the  SQS method~\cite{deOliveira2005} (section~\ref{sub:sqs}), which constrains the sampling to active configurations, and the RBC~\cite{Sander2016a} (section~\ref{sub:rbc}), in which the system returns to the pre-absorbing configuration  when absorbing one is visited. In the case of complex networks, which are highly heterogeneous, the localization on subextensive regions imposes further difficulties in the analysis of QS states~\cite{Mata2015,Silva2021}. While, on the one hand, the SQS is most general and able to capture localized phases of epidemics on networks~\cite{Mata2015,deArruda2017,Cota2016}, it has high computational and algorithmic complexity. On the other hand, RBC  is simpler but may distort  localized phases detected with SQS.

In order to soften these drawbacks, we introduced the  method called RAT (reactivation per activation time) intended to succeed where the SQS does but preserving as much as possible the algorithmic simplicity  and lower computational complexity of RBC. The method consists of reactivating vertices proportionally to the amount of time they were active along the whole history of the simulations. In the case of epidemic models, the activity  is defined as the total time that a vertex remains infected. This method can easily be generalized to other dynamical processes.

The RAT method was applied to the SIS model on a  variety of complex networks, which are characterized by strong localization effects~\cite{deArruda2017,Silva2021,Mata2015}  and compared with the RBC and SQS methods. We report that all methods provide the same epidemic threshold on multiplex and uncorrelated scale-free networks (degree exponent $\gamma<3$) but the  finite-size scaling of  the critical epidemic prevalence (density of active vertices at the threshold) of RBC differs of both SQS and RAT that match each other. Considering  random networks with outliers in the degree distribution, we report that RAT is able to capture the same localized activity as does the SQS, particularly in the cases where the RBC method deviates substantially from SQS. Finally,  simulations on real networks with degree correlations and very complex localization patterns for SIS dynamics support the equivalence of SQS and RAT while RBC distorts the  localized activation at low infection rates. Regarding the computational complexity, we observed that the RAM load of RAT is highly reduced compared to SQS while the CPU times near to the transition point are slightly smaller. Also, it worths remarking that RAT is algorithmically  simpler  than SQS.

Although the RAT method was thoroughly tested  for binary dynamics (only two states) in this work, it can be extended to other cases by introducing some procedures that depend  on  the model's details. For example, in the susceptible-infected-recovered-susceptible (SIRS) dynamics on networks~\cite{Ferreira2016a}, a modification of the SIS in which a healed vertex becomes immune (state R) to contagion and returns to the susceptible state with some fixed rate, the reactivation must be done immediately after the last infected vertex is healed, despite the rule still permits the spontaneous transitions R$\mapsto$S. As in the binary dynamics, the reactivation is done according to Eqs.~\eqref{eq:Ta} and \eqref{eq:na}. However, instead of introducing active vertices in a configuration in which all vertices are susceptible, one has to  use the state reached just after healing the last infected vertex. We expect that the RAT method will be extended to QS analyses of  many other processes. An interesting future prospect is to consider more complicated dynamics  with infinitely many absorbing states such as sandpile models~\cite{Manna1991,Dickman2001} and pair-contact process~\cite{Lubeck2002a}
and systems with bistability which undergo discontinuous transitions~\cite{deOliveira2015}.

\medskip

\begin{acknowledgments}
	We thank M. M de Oliveira, D. H. Silva and W. Cota for comments and suggestions.
This work was partially supported by the Brazilian agencies \textit{Conselho
	Nacional de Desenvolvimento Científico e Tecnológico}- CNPq (Grants no.
430768/2018-4 and 311183/2019-0) and \textit{Fundação de Amparo à Pesquisa do
	Estado de Minas Gerais} - FAPEMIG (Grant no. APQ-02393-18). This study was
financed in part by the \textit{Coordenação de Aperfeiçoamento de Pessoal de
	Nível Superior} (CAPES) - Brasil  - Finance Code 001.
\end{acknowledgments}

%

\end{document}